\documentclass[aps,floats,showpacs,epsf,twocolumn]{revtex4}
\usepackage{amsfonts}
\usepackage{graphics}
\usepackage{graphicx}
\usepackage{color}
\begin{document}

\title{Theory of unconventional quantum Hall effect in strained graphene}

\author{Bitan Roy$^{1}$, Zi-Xiang Hu$^{2,3}$, Kun Yang$^{1}$}

\affiliation{$^{1}$National High Magnetic Field Laboratory, Florida State
University, FL 32306, USA \\
$^{2}$Department of Physics, Chongqing University,
Chongqing 400044,  China \\
$^{3}$ Department of Electrical Engineering, Princeton University, Princeton, New Jersey 08544, USA}

\date{\today}

\begin{abstract}
We show through both theoretical arguments and numerical calculations that graphene discerns an unconventional sequence of quantized Hall conductivity, when subject to both magnetic fields (B) and strain. The latter produces time-reversal symmetric pseudo/axial magnetic fields (b). The single-electron spectrum is composed of two interpenetrating sets of Landau levels (LLs), located at $\pm \sqrt{2 n |b \pm B|}$, $n=0, 1, 2, \cdots$. For $b>B$, these two sets of LLs have opposite \emph{chiralities}, resulting in {\em oscillating} Hall conductivity between $0$ and $\mp 2 e^2/h$ in electron and hole doped system, respectively, as the chemical potential is tuned in the vicinity of the neutrality point. The electron-electron interactions stabilize various correlated ground states, e.g., spin-polarized, quantum spin-Hall insulators at and near the neutrality point, and possibly the anomalous Hall insulating phase at incommensurate filling $\sim B$. Such broken-symmetry ground states have similarities as well as significant differences from their counterparts in the absence of strain. For realistic strength of magnetic fields and interactions, we present scaling of the interaction-induced gap for various Hall states within the zeroth Landau level.
\end{abstract}


\maketitle

\vspace{10pt}

Successful fabrication of two-dimensional electron gas, e.g, galium-arsenide (GaAs) heterostructure, provided a unique opportunity to observe a novel aspect of low-dimensional electronic systems, quantization of Hall conductivity ($\sigma_{xy}$). At weaker magnetic fields ($\sim 1$ T), even the low-mobility samples discern quantized plateaus of $\sigma_{xy}$ at various integers of $e^2/h$. This phenomena is referred to as the \emph{integer quantum Hall effect} (IQHE)\cite{IQHE}. A rather more striking observation is the plateaus of that quantity at various, for example 1/3, fractions of $e^2/h$, in improved samples, however at stronger fields ($\sim 10$ T)\cite{FQHE}. Whereas the IQHE arises from the free motion of fermions in magnetic fields \cite{prange}, its fractional version necessarily requires strong electron-electron interactions to develop the mobility gap within a partially filled Landau level(LL) \cite{laughlin}.

Integer quantization of $\sigma_{xy}$ occurs when the chemical potential ($\mu$) lies within a mobility gap, filled by localized states, separated by two extended conducting edge modes carrying the quantized Hall current \cite{halperin}. As the magnetic field (B) is reduced, more and more extended states, at well-separated energies, get occupied. Total Hall current, the \emph{algebraic} sum of it carried by each of the edge modes, then encounters quantized increment, due to identical \emph{chirality} of all the edge states \cite{laughlinthoughtexpt}.

Besides the GaAs heterostructure, the new generation two-dimensional electronic system, \emph{graphene}, discerns a sequence of Hall plateaus at fillings $\nu= \pm 4 (n+\frac{1}{2})$, subject to relatively low fields \cite{geim}, while additional plateaus, for example at $\nu=0,\pm1,\pm4$, show up as the field is enhanced \cite{kim,barlasreview12}. Otherwise, all the LLs support the current-carrying states with identical chirality, as in GaAs \cite{graphenechirality}. Moreover, due to its mechanical flexibility under strain, graphene may experience yet another \emph{effective} magnetic field, resulting from deliberate bulging \cite{levy}. Such strain-induced pseudo/axial magnetic field (b) preserves the time-reversal symmetry (TRS), and points in opposite directions at two inequivalent Dirac points, suitably chosen here at $\vec{K}=(1,1/\sqrt{3}) (2\pi/a \sqrt{3})$ and $-\vec{K}$ \cite{semenoff}. Therefore, subject to strain as well as an external magnetic field, one can expose the gapless Dirac quasi-particles, near two Dirac points with different effective fields, $|B\pm b|$, possibly pointing in {\em opposite} directions, respectively. Hence, an interplay of these two gauge fields, concomitantly an unconventional quantization of the Hall conductivity, can be realized in graphene.

It is perhaps worth considering the Hall response of this system when $B>b (\neq 0)$ first\cite{fieldstrength}. The spectrum of non-interacting Dirac quasi-particles is then composed of two inter-penetrating sets of LLs at well separated energies $\pm \sqrt{2 n (B\pm b)}$, with degeneracies $(B\pm b)/2 \pi$ per unit area, and all the LLs experience the effective orbital magnetic fields in the same direction. Hence, every current-carrying state has identical chirality. Consequently, as the chemical potential sweeps through various LLs, the total Hall current adds up, and the quantization of $\sigma_{xy}$ is expected to occur at all integers of $e^2/h$. However, the plateaus appear at incommensurate fillings, due to distinct degeneracies of the LLs\cite{bitanoddQHE}.

\begin{figure}[htb]
\includegraphics[width=4.1cm,height=3.1cm]{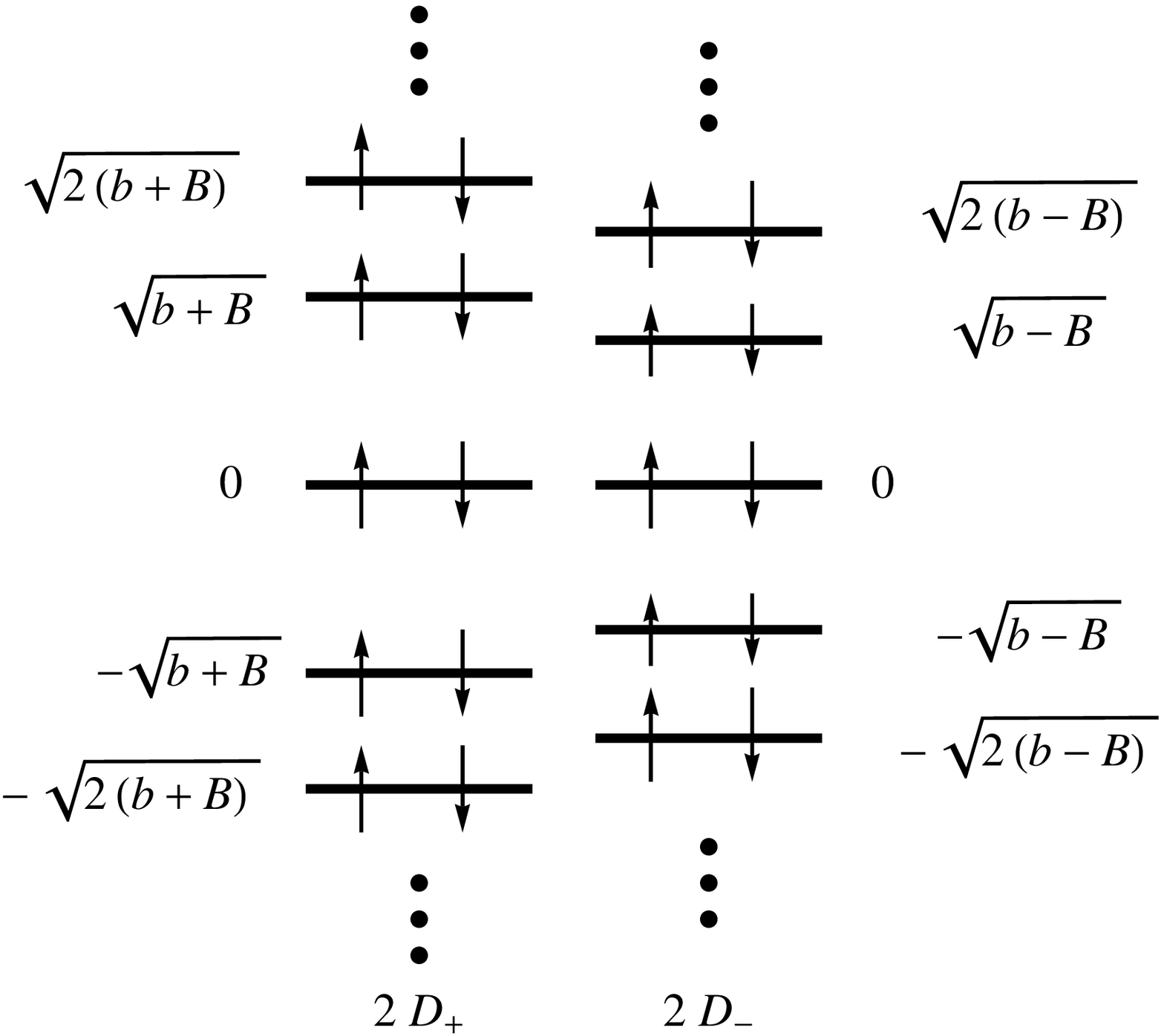}
\includegraphics[width=4.0cm,height=3.1cm]{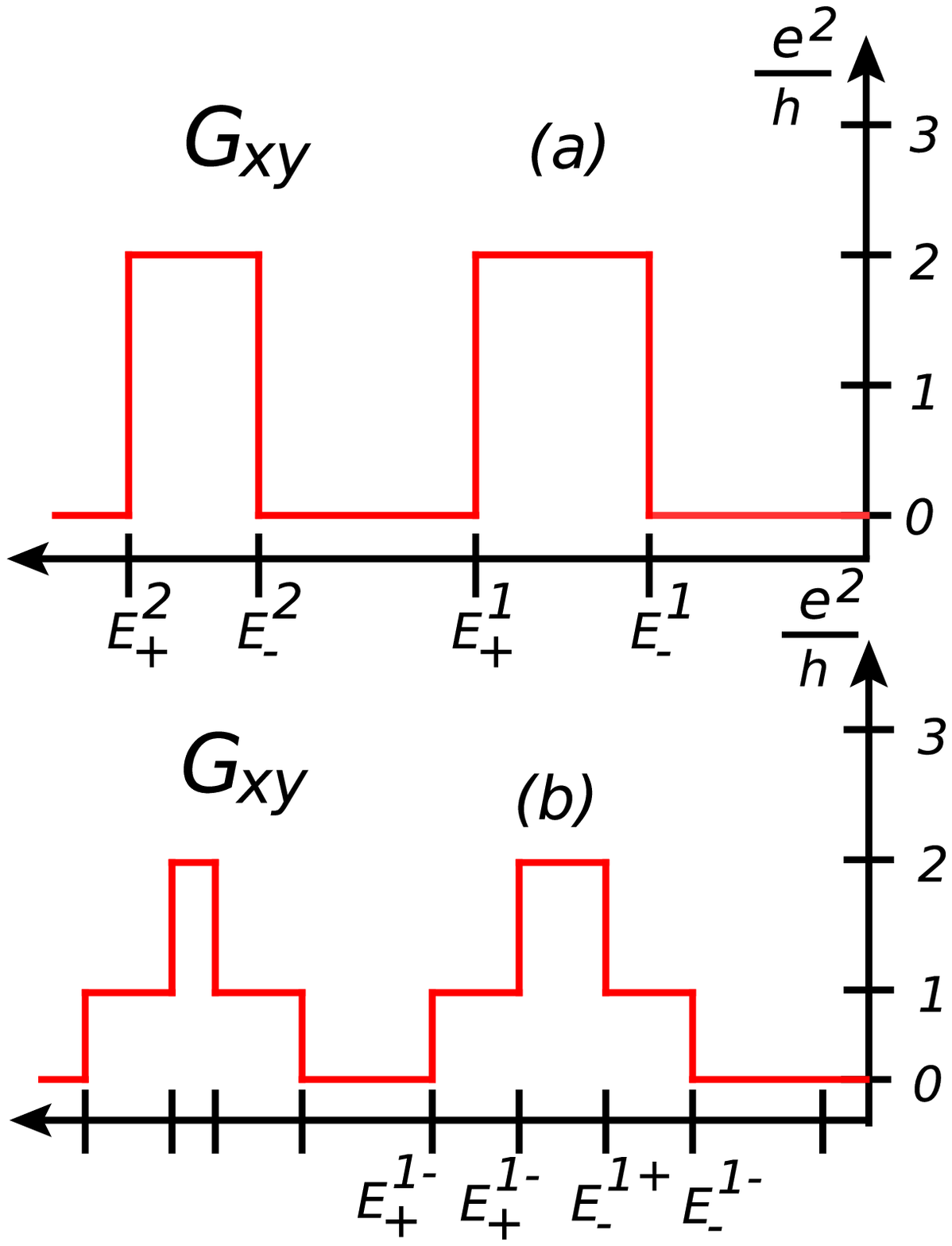}
\includegraphics[width=8.75cm,height=3.75cm]{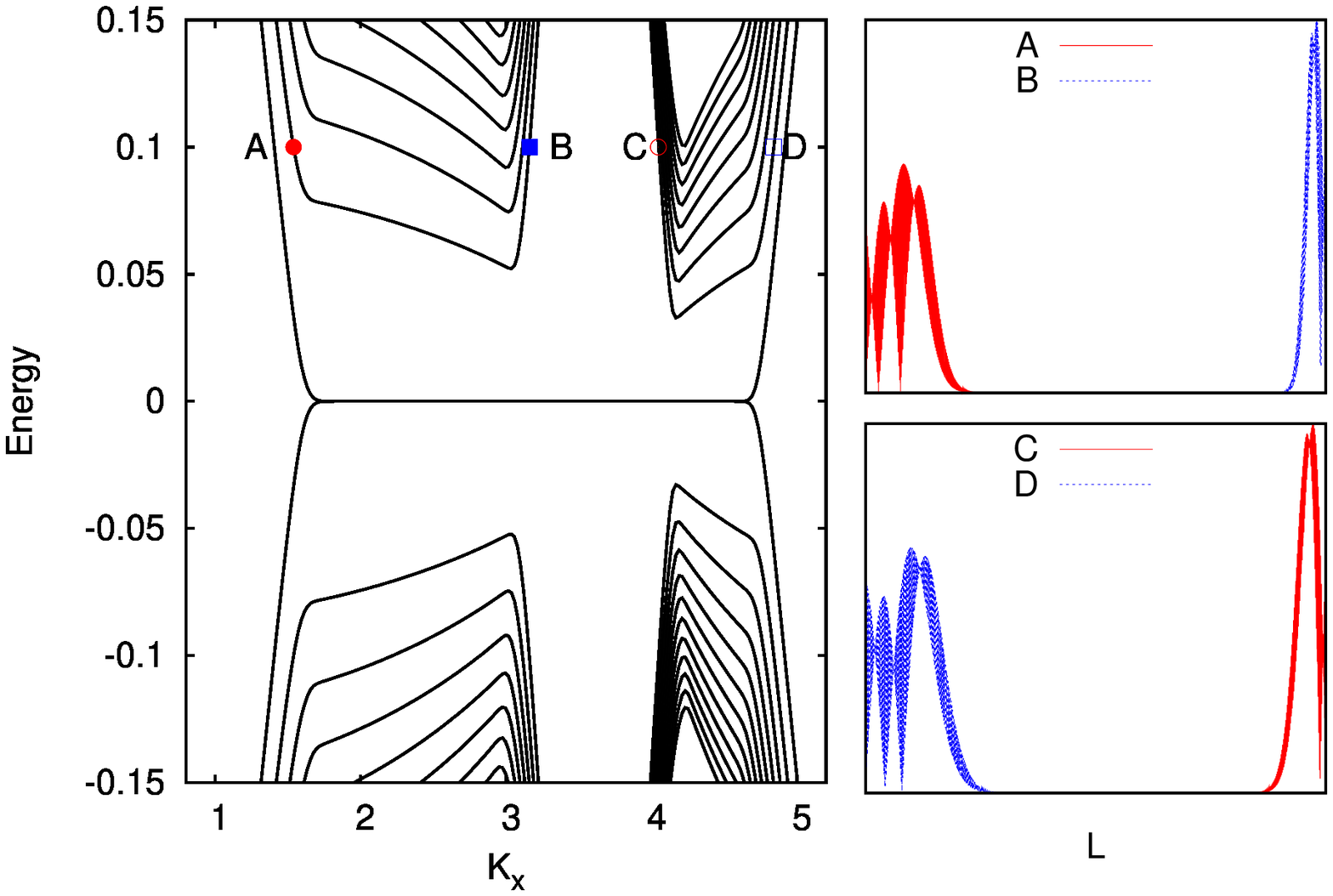}
\caption[] {(Color online) Upper row:: Left: Spin-degenerate interpenetrating LLs of $H_D[A,a]$. Here we have shown the LLs for $n=0,\pm 1,\pm 2$ only. Two LLs have the degeneracies $2 D_\pm=(b \pm B)$ per unit area. Right: Schematic variation of Hall conductances $G_{xy}$ in a hole-doped graphene, without (a) and with (b) Zeeman splitting ($\Delta_z$). Here, $E^n_{\sigma}=\sqrt{2 n (b + \sigma B)}$, $E^{n \alpha}_\sigma=E^n_{\sigma}+\alpha \Delta_z$ with $\sigma, \alpha=\pm$. In the electron-doped system, $G_{xy}$ changes sign. We only show the spin splitting of $n=1$ LL. Splitting of $n=2$ LL is identical. Lower row:: The energy spectrum (left) and wave functions (WFs) (right) for a strained graphene in magnetic field when $B < b$. WFs localized on one edge live on the opposite side at two valleys, explicitly, A(B) and D (C) are localized on left (right) edge, therefore carrying opposite chirality.}\label{freeLL}
\end{figure}

A rather more interesting situation arises when $b>B$. For $B=0$, the pseudo Dirac LLs, placed at $\pm \sqrt{2 n b}$ \cite{herbutpseudo, herbut-roy-pseudo, ghaemi, abanin}, near two valleys have opposite chirality, henceforth the TRS is preserved. As long as $b>B>0$, two inequivalent sets of LLs, now located at $\sqrt{2 n (b \pm B)}$ (Fig. \ref{freeLL}, left column), with respective degeneracies $(b \pm B)/2 \pi$ per unit area, continue to enjoy opposite chirality (Fig. \ref{freeLL}, lower right column). Consequently, as the chemical potential starts to deviate from the charge neutrality point (CNP), the Hall conductivity is restricted within $\pm 2 e^2/h$ (when more LLs near one valley are filled) and $0$ (when both valleys are equally populated); see Fig. \ref{freeLL}(upper right column). The $\pm$ sign corresponds to hole and electron doped systems, respectively, and note it is {\em opposite} to what one has in the absence of the strain-induced field, $b$.

Even though the Hall conductivity stays bounded, as the chemical potential is enhanced, more and more current-carrying edge states with opposite chirality get filled. In the absence of back scattering that equilibrates these counter-propagating edge modes, the two-terminal conductance $G_{xx}$ is expected to increase monotonically. However, the lack of equilibration ruins the quantization of Hall conductance in a four-terminal measurement\cite{Kane-Fisher}. But in reality, there is always back scattering between counter-propagating edge modes that live along the same edge; this not only equilibrates these modes but also localizes them, except for the two additional modes associated with the occupied extra LLs. As a result both $\sigma_{xy}$ and $G_{xx}$ are quantized at the same value.

The oscillatory sequence of $\sigma_{xy}$ is strictly true only in the vicinity of the CNP. The spacing of the Dirac LL decreases with the LL index (n), and the effective magnetic field for two sets of LLs are different. Hence, far away from the CNP, LL crossing is unavoidable, and one may see quantized plateaus of $\sigma_{xy}$ at $3 e^2/h$ or higher. If $B \ll b$, the LL crossing occurs for $n \gg 1$. Assuming that the chemical potential is not too far from the Dirac points, one can then safely neglect the LL crossing. 

To compute the LL spectrum, we here construct an 8-component Dirac spinor $\Psi=\left(\Psi_\uparrow, \Psi_\downarrow \right)^\top$, where $\Psi^\top_\sigma=[ u^\dagger _{\sigma} (\vec{K}+\vec{q}), v^\dagger_{\sigma} (\vec{K}+\vec{q}), u^\dagger _{\sigma} (- \vec{K}+\vec{q}), v^\dagger _{\sigma}(-\vec{K}+\vec{q}) ]$, with $\sigma=\uparrow,\downarrow$ as electrons spin projection along the $z-$direction. The orbital effects of the real (B) and pseudo (b) magnetic fields can be captured by the Hamiltonian \cite{bitanoddQHE, herbutpseudo,herbut-roy-pseudo}
\begin{equation}
H_D \left[ A, a \right]=I_2 \otimes i\gamma_0 \gamma_i \left( \hat{q}_i -A_i -i \gamma_{3} \gamma_5 a_i \right),
\label{DiracHamilmagneticfields}
\end{equation}
where $B (b)=\epsilon_{3 i j} \partial_i A(a)_j$. The gamma matrices are $\gamma_0 = I_2 \otimes \sigma_3$, $\gamma_1= \sigma_3 \otimes \sigma_2$, $\gamma_2= I_2 \otimes \sigma_1$, $\gamma_3= \sigma_1 \otimes \sigma_2$, $\gamma_5= \sigma_2 \otimes \sigma_2$ \cite{herbut-juricic-roy}. The spectrum of $H_D[A,a]$ is composed of two sets of \emph{interpenetrating} LLs at energies $\pm \sqrt{2 n (b+ \alpha B)}$, with respective degeneracies $\Omega (b + \alpha B)/2 \pi$ for $\alpha=\pm$, shown in Fig. \ref{freeLL} \cite{supplementary}. Here $n=0,1,2,\cdots$ and $\Omega$ is the area of the strained graphene sample. With $b>B$ states within the zeroth LL (ZLL) are localized on only one sub-lattice, say A for example, while they reside on complimentary sub-lattices near two Dirac points if $B>b$ \cite{bitanoddQHE}. For each spin flavor, there exists $(b \pm B) \Omega$ states at precise zero energy per unit area, guaranteed by an ``index theorem" \cite{indextheorem, roy-inhomogeneous}, respectively near the Dirac points at $\pm \vec{K}$. However, the valley degeneracy for all the LLs at finite energies is removed, as they are exposed to different \emph{effective} magnetic fields.

Let us first register the Hall response of the non-interacting system, see Fig. \ref{freeLL}(a). For $\mu=0$, the ZLL, containing $4 \Omega b$ states, is half filled. Then a particle-hole symmetry of the spectrum, generated by $I_2 \otimes \gamma_0$ for example\cite{PHsymm}, guarantees that $\sigma_{xy}=0$. Even when $0< \mu < \sqrt{b-B}$, $\sigma_{xy}$ remains pinned at zero, which can be confirmed upon subscribing to the St\v{r}eda formula \cite{streda} for the Hall conductivity, read as $\sigma_{xy}=\left( \frac{\partial N}{ \partial B}\right)_\mu$, in the natural units $e=c=1$. $N$ is the electronic density in the bulk below the chemical potential. The derivative with respect to $B$ is taken at fixed $\mu$, measured from the half-filled band. In order to place the chemical potential such that $0 < \mu < \sqrt{b-B}$, one needs to fill $N=2 \Omega b$ (independent of $B$) states from the CNP, yielding a \emph{zero} Hall conductivity. On the other hand, if $\sqrt{b-B}< \mu < \sqrt{b+B}$, $\delta N= - 2 \delta B$ and hence $\sigma_{xy}=-2$. The factor 2 counts the spin degeneracy of the LL. Upon further doping when $\sqrt{b+B}< \mu <\sqrt{2 (b-B)}$, the Hall conductivity returns to \emph{zero}. Hence, with odd (even) number (modulo 2 due to spin) of LLs below the chemical potential, one gets $\sigma_{xy}=-2 (0)$, as long as there is no LL crossing. The origin of such oscillating Hall conductivity is the following. Two sets of LLs near $\pm {K}$ experience effective magnetic fields $(b \pm B)>0$, but point in opposite directions. So, the current-carrying states of these two LLs have opposite \emph{chirality}. When an odd number (modulo 2) of LLs above the CNP are filled, $\sigma_{xy}=-2$, since there is imbalance in the occupation of the LLs near two Dirac points. With an even number (modulo 2) of filled LL, the Hall currents from the LLs near $\pm \vec{K}$ exactly cancels each other, giving $\sigma_{xy}=0$. The Hall conductivity oscillates between $0$ and $+2 \; e^2/h$, when $\mu<0$ (hole doping).
\begin{figure}[htb]
\includegraphics[width=4.1cm,height=3.5cm]{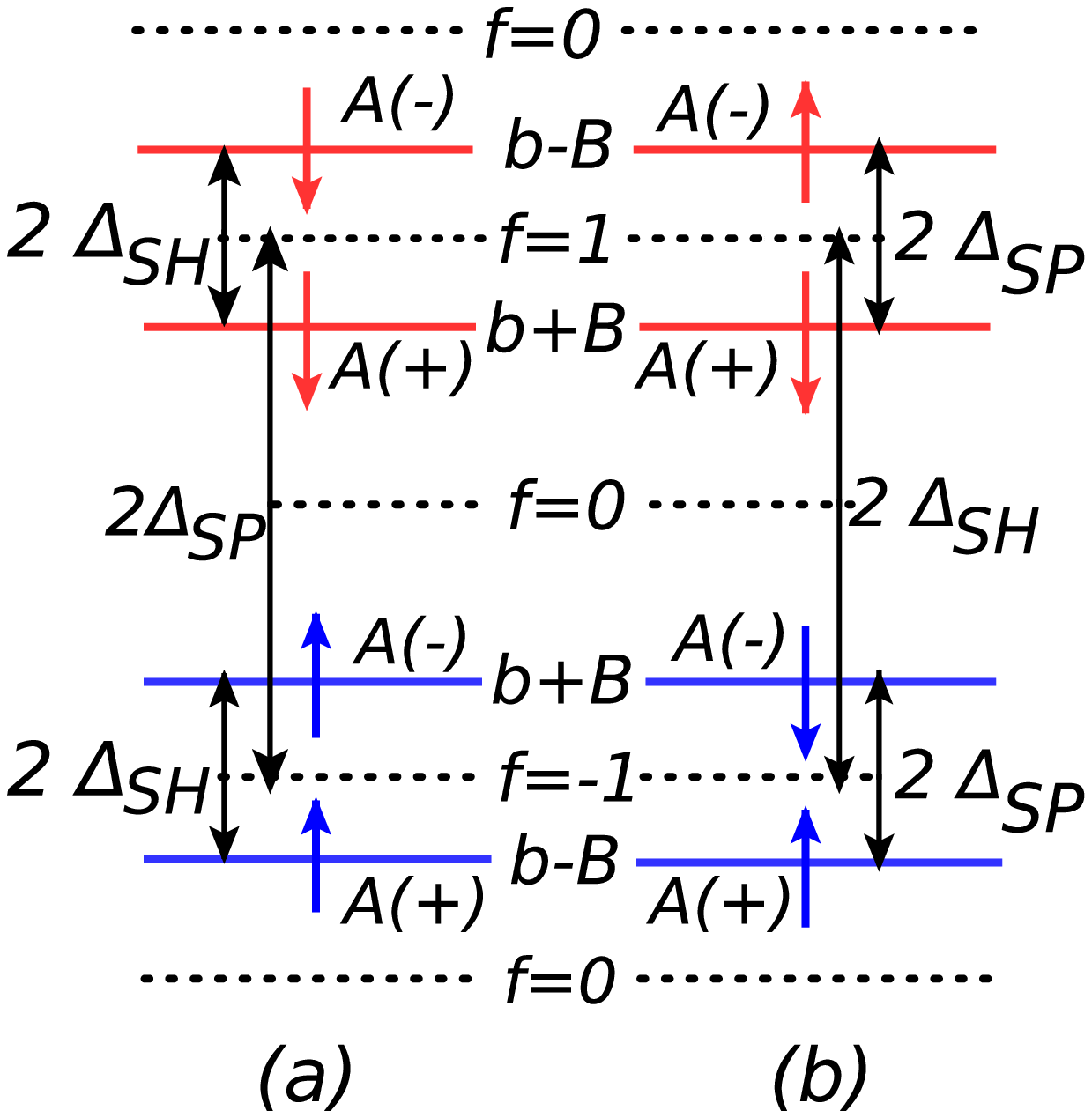}
\includegraphics[width=4.25cm,height=3.5cm]{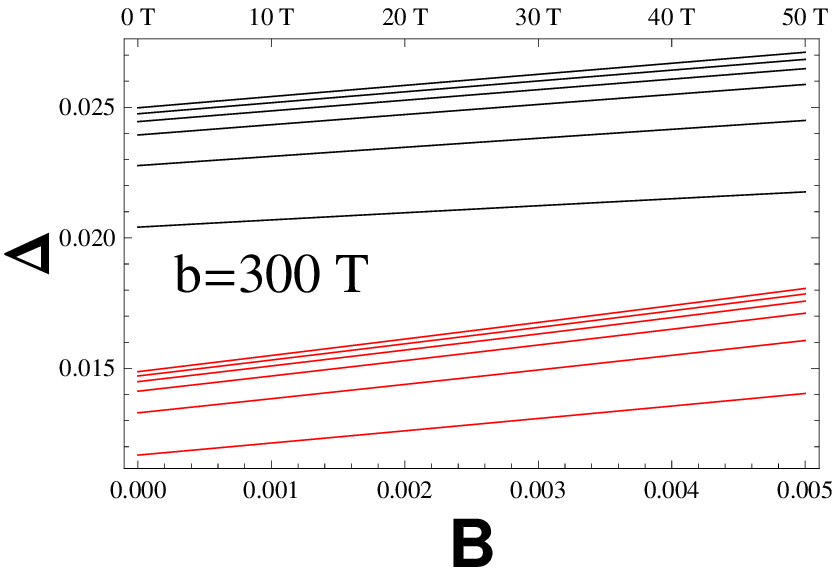}
\caption[] {(Color online) Left: Two possible [(a),(b)] interaction-driven splittings of the ZLL. Dotted lines correspond to requisite location of the chemical potential for $\sigma_{xy}=f e^2/h$. $\pm$ corresponds to the ZLL states localized near the valley at $\pm \vec{K}$. Right: Dimensionless activation gap ($\Delta/\Lambda$) for $\sigma_{xy}=0, \pm e^2/h$ (black, red) Hall states when $b=300$ T and $0$ T $< B < 50$ T, for $\delta=0, 0.03,0.07,0.14,0.3,0.7$ (from top to bottom), assuming $\Lambda \sim 1/2.5 \mathring{A}$, the ultra-violet cutoff over which the dispersion is approximately linear. $g_c$ is the zero-field criticality. Lower x axis denotes $B/\Lambda^2$ (dimensionless).}\label{ZLLQHE}
\end{figure}

The chiral nature of the edge modes in the presence of strain and magnetic field can also be seen in a finite honeycomb lattice, with the zigzag edge. The orbital effect of the real and axial magnetic field can respectively be captured by, attaching a Peierls phase, $t_{ij} \rightarrow t_{ij} \exp{(2 \pi i \frac{e}{hc}\int_i^j \vec{A}\cdot d\vec{l}})$, and introducing local modification, to the nearest-neighbor (NN) hopping amplitudes. We here modify the hopping only along one of the three bonds, oriented orthogonal to the zigzag edge \cite{supplementary, chang}. Such simple deformation yields slightly inhomogeneous Fermi velocities and thus LL energies. Nevertheless, one can still observe the peculiarities of the edge modes arising from the time-reversal invariance. It is evident from Fig. \ref{freeLL}(lower row) that the chiralities of two states localized near one zigzag edge are opposite if $b>B$, similar to when $B=0$, but $b \neq 0$.\cite{supplementary} LLs near two Dirac points appear at different energies. Therefore, as one changes the chemical potential the Hall conductivity keeps oscillating between $\pm e^2/h$ and $0$ (we consider spinless fermions). If on the other hand, the real magnetic gets stronger, the edge modes near two Dirac points share identical chiralities, and $\sigma_{xy}$ changes monotonically.\cite{supplementary, bitanoddQHE} An interesting possibility is $B=b$. Only one of the Dirac points is then exposed to finite magnetic field, yielding plateaus of $\sigma_{xy}$ at $\nu=2 (n+\frac{1}{2})$, while the other one remains semi-metallic, contributing to $\sigma_{xx}$ simultaneously.\cite{guinea-low, supplementary}

The Zeeman splitting ($\Delta_z$) lifts the spin degeneracy from all the LLs, $\sqrt{n (b\pm B)} \rightarrow \sqrt{n (b\pm B)} \pm \Delta_z$. The Zeeman gap scales as $\Delta_z \sim B$(Tesla) K. Hence, it cannot cause LL crossing near the CNP. The Hall conductivity remains pinned to \emph{zero}, when $\mu=0$ due to the particle-hole symmetry, generated by $\sigma_1 \otimes \gamma_0$.\cite{PHsymm} It remains so even when the chemical potential lies in between the Zeeman shifted  ZLL at $\Delta_z$, and $\sqrt{b-B}-\Delta_z$, since then $\delta N=2 b \Omega$ in the St\v{r}eda formula. Otherwise, for $\sqrt{b+ \sigma B}+ \sigma \Delta_z < \mu < \sqrt{b+ \sigma B}-\sigma \Delta_z$, $\sigma_{xy}=- \left( \frac{3+\sigma}{2}\right) e^2/h$, where $\sigma=\pm 1$. Therefore, the Zeeman splitting introduces additional Hall plateaus at $ \mp e^2/h$, in electron and hole doped systems, respectively; see Fig. \ref{freeLL}(b).

A strain induced charge density wave order (not spontaneously generated) always persists within the ZLL, since all these states are localized on one sub-lattice. This configuration is a natural ground state for the residual NN Coulomb repulsion. Two valleys at $\pm \vec{K}$ hosts $\Omega (b \pm B)$ states, hence a ``valley polarized" \emph{anomalous Hall insulator} cannot develop at the CNP. It may however be realized at incommensurate filling, $\nu \propto B$ about the neutrality point. A spin Hall (SH) order, $\Delta_{SH}=\langle \Psi^\dagger_\sigma \vec{\sigma} \otimes i \gamma_1 \gamma_2 \Psi_\sigma \rangle$\cite{haldane, kekuleSC}, corresponding to a spin-triplet, intra-sublattice circulating currents, can however develop upon occupying two valleys with opposite spin. It also carries a finite \emph{ferromagnetic} moment $\propto B$, the difference of LL degeneracies. Since the entire ZLL is localized on one sub-lattice, a ferromagnetic (FM) order is tied with an anti-ferromagnetic (AF) order. Yet another state, namely a spin polarized (SP) state can also be realized at the CNP. It carries FM ($\Delta_{FM}=\langle \Psi^\dagger_\sigma \vec{\sigma} \otimes I_4 \Psi_\sigma \rangle$) and AF ($\Delta_{AF}=\langle \Psi^\dagger_\sigma \vec{\sigma} \otimes \gamma_0 \Psi_\sigma \rangle$) orders, simultaneously \cite{herbut-roy-pseudo}. The Zeeman coupling locks the spin of the SP state along the direction of the real field ($B$), and gives $\Delta_{FM} \neq 0$. However, the on-site Hubbard interaction (U), possibly the strongest interaction in graphene \cite{katsnelson}, favors an AF order in the vicinity of the CNP. The second-neighbor repulsion ($V_2$) favors the SH state\cite{raghu}. The spin polarized state can also be realized even when $B=0$ \cite{ghaemi}, which has been identified as pure ferromagnetic state in Ref. \cite{ghaemi}. The AF/SH order parameter anti-commutes with $H_D[A,a]$. Hence, apart from splitting the ZLL, they optimally lower the energy of the filled Dirac sea by shifting all the LLs at finite energies, $\pm \sqrt{2 n (b \pm B)} \to \pm \sqrt{2 n (b \pm B)+\Delta^2_{AF/SH}}$. The spin-polarized gap within the ZLL is $\Delta_{SP}=\Delta_{AF}+\Delta_{FM} \sim U$, whereas $\Delta_{SH} \sim V_2$, to the leading order.\cite{bitan-scaling} Though such insulations in pristine graphene can only take place for sufficiently strong repulsive interactions,\cite{herbut-juricic-roy} the existence of macroscopically degenerate LLs permits such ordering even when the interactions are infinitesimal, in the presence of magnetic fields.\cite{catalysis, herbutpseudo, herbut-QHE, bitanoddQHE} Next we study the interplay of these orders.

For small Zeeman coupling the ground-state energy per unit area at half filling with AF ($\vec{N}$) and SH ($\vec{C}$) orders is
\begin{equation}
E [\vec{N},\vec{C}]=\frac{\vec{N}^2}{4 g_a}+\frac{\vec{C}^2}{4 g_c}+ E_0 [\vec{N},\vec{C}],
\end{equation}
where $g_{a(c)} \sim U (V_2)$. $E_0[\vec{N},\vec{C}]$ is the ground-state energy per unit area of the effective single-particle Hamiltonian
\begin{equation}
H_{HF}=H_D[A,a]- \left( \vec{N}\cdot \vec{\sigma} \right) \otimes \gamma_0 -\left(\vec{C} \cdot \vec{\sigma} \right) \otimes i\gamma_1 \gamma_2.
\end{equation}
With negligible Zeeman coupling a spin-anisotropy can be neglected and one can take $\vec{N}(\vec{C})=\left( N(C),0,0\right)$, for simplicity. The spectrum of $H_{HF}$ is composed of LLs at energies $\pm e_{n,\alpha}$, with degeneracies $\Omega (b+\alpha B)/4\pi$, where $e_{n,\alpha}= [2 n (b+\alpha B)+ (N + \alpha C)^2]^{1/2}$, and $\alpha=\pm$. The ground state of $H_{HF}$ at half filling has all the states with negative energies filled, while the rest are empty. Minimization of $E [N,C]$, with respect to $N$ and $C$, yields \emph{two} coupled gap equations, which for $N>C$ read as
\begin{equation}
\frac{\pi^{3/2}}{2 g_i}= \frac{\xi_i}{X_i}+ \sum^{i\neq j}_{n \geq 1, \alpha=\pm} \left(\frac{b + \alpha B}{2 e_{n,\alpha}} +\frac{X_j}{ X_i} \alpha \frac{b + \alpha B}{2 e_{n,\alpha}} \right)
\end{equation}
for $i,j=1,2$, where $g=(g_a,g_c), X=(N,C), \xi=(b,B)$. The ultra-violet divergence in the first term of the gap equations is independent of AF or SH orders. The cut-off ($\Lambda$) independence of the physical observable gap, then demands $g_a \equiv g_c$ for both AF and SH orders to be finite  simultaneously. Since in graphene $U > V_2$, possibly a spin polarized state ($N\equiv \Delta_{AF} \neq 0, C\equiv \Delta_{QSH}=0$) is formed at the CNP. Even though, with $b>B$, there exists a series of $\sigma_{xy}=0$ plateau, only the one near $\mu=0$ bears an AF order, while the rest arises from lack of ``valley reflection symmetry". Placing the chemical potential close to the first excited state at $\pm \Delta_{SP}$, a spin Hall order develops an additional incompressible Hall state, leading to $\sigma_{xy}=\pm e^2/h$, see Fig. 2(a). If on the other hand, $V_2 > U$, yielding $C>N$, the splitting of the ZLL gets reversed, see Fig. 2(b).

Minimizing the ground state energy, one can find the gap equation for $\sigma_{xy}=0$ Hall state near the CNP. For fixed axial magnetic field ($b=300 T$), the interaction induced gap at the neutrality point increases linearly with the real magnetic field, when $0$ T $< B < 50$ T, (Fig. 2, right column). Scaling of the gap is insensitive to the exact nature of the order parameter. The activation gap for the $\sigma_{xy}=\pm e^2/h$ state within the ZLL is smaller than, but similar to that for the $\sigma_{xy}=0$. Such hierarchy comes solely from the ZLL. Exactly half of the ZLL contributes to the gap for the $\sigma_{xy}=0$ state, whereas fewer states from the ZLL contribute to the gap for the $\sigma_{xy}=\pm 1$ Hall state.\cite{supplementary} Otherwise, activation gaps for both the Hall states scale sub-linearly with the interaction, $\delta=(g \Lambda)^{-1}-(g_* \Lambda)^{-1}$, where $g_*$ is zero-field criticality for insulation. If the magnetic fields become \emph{inhomogeneous}, the LLs at finite energies disappear, giving rise to a continuous spectrum, though the ZLL, protected by the ``index theorem", stays unaffected. Therefore, interaction-induced gap formation occurs even when the fields are non-uniform. However, the gaps then closely follow the profile of the magnetic fields.\cite{roy-inhomogeneous} With \emph{weak} inhomogeneous fields, the quantization of $\sigma_{xy}$ is expected to survive.

In the absence of the axial field or even when $B>b$, the states within the ZLL, localized near two valleys, live on complementary sub-lattices \cite{herbut-QHE, bitanoddQHE}. Therefore, a conventional AF order develops by filling up states on two sublattices with opposite spin projections. However, the staggered spin moments on two sub-lattices are of different magnitudes. Therefore, one may argue such a correlated phase as \emph{ferrimagnetic} as well \cite{bitanoddQHE}.

In experiment \cite{levy}, the \emph{uniform} axial field is localized only in a certain region of the sample. Particles circling that region pick up an axial Aharonov-Bohm phase (ABP), only if they travel through the strained region, since the axial gauge potential is proportional to strain. It is identical for the trajectories in opposite directions, whereas the ABPs due to the real magnetic field are of opposite sign for these trajectories.\cite{vozmediano-nat-phys} Consequently, the trajectories with opposite circulation, acquires different effective ABPs, namely, the sum and difference of it due to two fields. In Hall conductivity measurement, the terminals need to be attached to the regions with at least finite strain, though $b$ can be zero. In \emph{molecular graphene}\cite{manoharan}, and strained graphene on Ru substrate \cite{castro-Ru-buble}, the axial field can possibly be realized in the entire sample. Hence the peculiar Hall conductivity, we propose here, may become easily observable in those systems.

To summarize, we here demonstrate the possible quantization of Hall conductivity ($\sigma_{xy}$) in strained graphene, subject to magnetic fields. We show that when the strain-induced pseudo magnetic field is stronger than the real one, Hall conductivity remains bounded between $0$ and $\mp 2 e^2/h$, in electron and hole doped graphene respectively. The Zeeman coupling introduces additional Hall plateaus at $\mp e^2/h$. Such quantization relies on sufficient backscattering among the counterpropagating edge modes, and is only true in the vicinity of the CNP, where LL crossing can safely be neglected. Depending on the relative strength of the finite-ranged components of the Coulomb interaction, various broken-symmetry phases can be realized within the ZLL. For example, on-site and next-neighbor repulsion respectively favors anti-ferromagnetic and spin Hall ground states. In contrast to the conventional situation, the antiferromagnetic order, in the strain dominated regime ($B \ll b$), is always tied with a ferromagnetic order. For fixed $b$, the many-body gaps inside the ZLL scale linearly with real magnetic field (B) and sublinearly with interaction ($\delta$), as long as $B \ll b$.

B.R. acknowledges the support by National Science Foundation Cooperative Agreement No. DMR-0654118, the State of Florida, and the U.S. Department of Energy. Z. X. H. is supported by NSFC No. 11274403 and DOE Grant No. de-sc0002140. K. Y. is supported by NSF Grant No. DMR-1004545. B. R. is grateful to Igor F. Herbut for many useful discussions. B.R. thanks \'{E}cole de Physique, Les Houches for hospitality during the summer school ``Strongly Interacting Quantum Systems Out of Equilibrium", where part of the manuscript was prepared.

\vspace*{3cm}

\begin{widetext}
\begin{center}
{\bf \large Supplementary material for ``Theory of unconventional quantum Hall effect in strained graphene"}
\end{center}
\end{widetext}

\vspace{10pt}
\section{Diagonalization of Dirac Hamiltonian in real and axial magnetic fields}

We here present one possible prescription to diagonalize the Dirac Hamiltonian in the presence of real and strain induced pseudo/axial magnetic fields. Recall, the low energy Hamiltonian in graphene, subject to strain and magnetic field reads as \cite{pseudorealdirac}
\begin{equation}
H_D \left[ A, a \right]=I_2 \otimes i\gamma_0 \gamma_i \left( \hat{q}_i -A_i -i \gamma_{3} \gamma_5 a_i \right).
\end{equation}
The axial vector potential $(a_i)$ is a member of general $SU(2)$ time reversal symmetric gauge potential
\begin{equation}
a_i=\gamma_3 a^3_i+\gamma_5 a^5_i+ i \gamma_3 \gamma_5 a^{35}_i.
\end{equation}
\begin{figure}
\includegraphics[width=7.5cm]{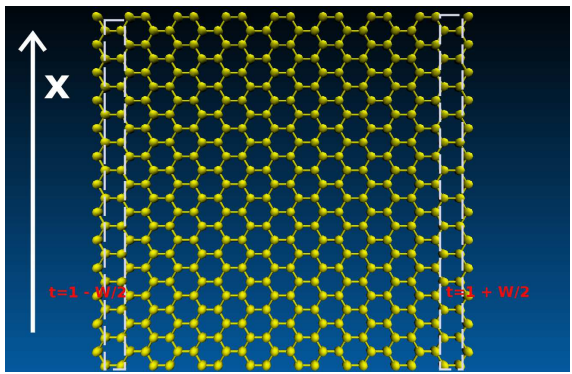}
\caption{\label{sketch} (Color online) Graphene with zigzag edges. The edge (along $x$ direction) is effectively infinite and the width in $y$ direction is finite. In the strained case, we suppose the hopping terms along $y$ direction has a gradient which ranges from $t = 1 - W/2$ to $t = 1 + W/2$ from one side to the other.}
\end{figure}
However, the first two terms break the translational symmetry, generated by $I_{tr}=I_2 \otimes i \gamma_3 \gamma_5$. If the deformation of the graphene flake is smooth enough one can safely neglect their contributions and only keep the $a^{35}_i$ component. Upon setting $a^3_i=a^5_i=0$, we set $a^{35}_i\equiv a_i$, for notational simplicity. Thereafter, the Dirac Hamiltonian is devoid of any valley mixing. One can then exchange the second and the third $2 \times 2$ blocks of $H_D [A,a]$ and cast it in a block diagonal form, say $H_+ \oplus H_-$, where 
\begin{equation}
H_{\pm}=\pm I_2 \otimes \sigma_1 (\hat{q}_1 \mp a_1 -A_1) + I_2 \otimes \sigma_2 (\hat{q}_2 \mp a_2 -A_2).
\end{equation}  
$H_\pm$ capture the effect of both the gauge potentials on the Dirac fermions in the vicinity of $\pm \vec{K}$ points, respectively. However, both $H_\pm$ are unitarily equivalent to the celebrated Dirac Hamiltonian, $H_D= i \gamma_0 \gamma_i (\hat{q}_i - A^{ef}_i)$, subject to effective magnetic fields, $b \pm B$ respectively. Explicitly, $H_+=U_1^\dagger H_D U_1$, with $U_1=I_2\oplus i\sigma_2$ and $A^{ef}_i=a_i+A_i$, while $H_-=U_2^\dagger H U_2$, with $U_2= i\sigma_2 \oplus I_2$ and $A^{ef}_i=a_i-A_i$ \cite{pseudorealdirac,igorunitary}. After expressing the original Hamiltonian $H_D[A,a]$ as direct sum of two inequivalent copies of the standard Dirac Hamiltonian subject to different effective magnetic fields, one can immediately diagonalize it, yielding the announced spectrum of inter-penetrating sets of Landau levels (LLs) \cite{LLDirac}.  
\begin{figure}
\includegraphics[width=4cm]{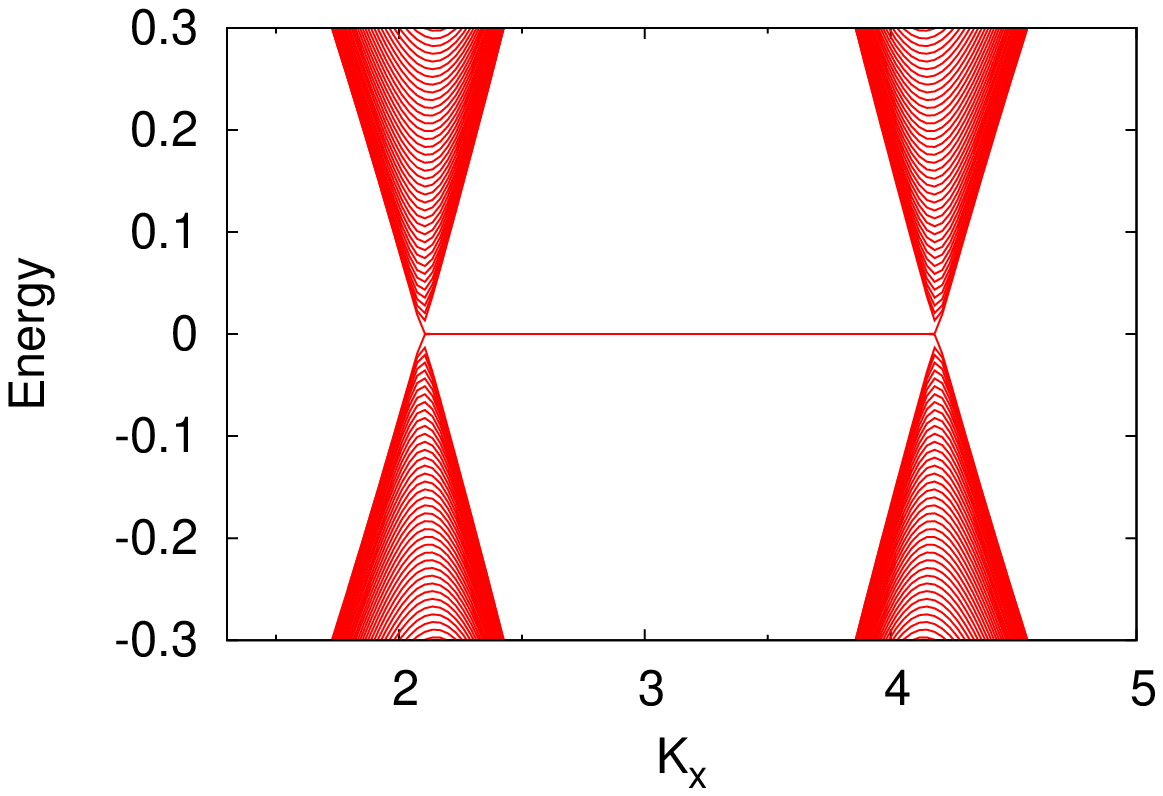}
\includegraphics[width=4cm]{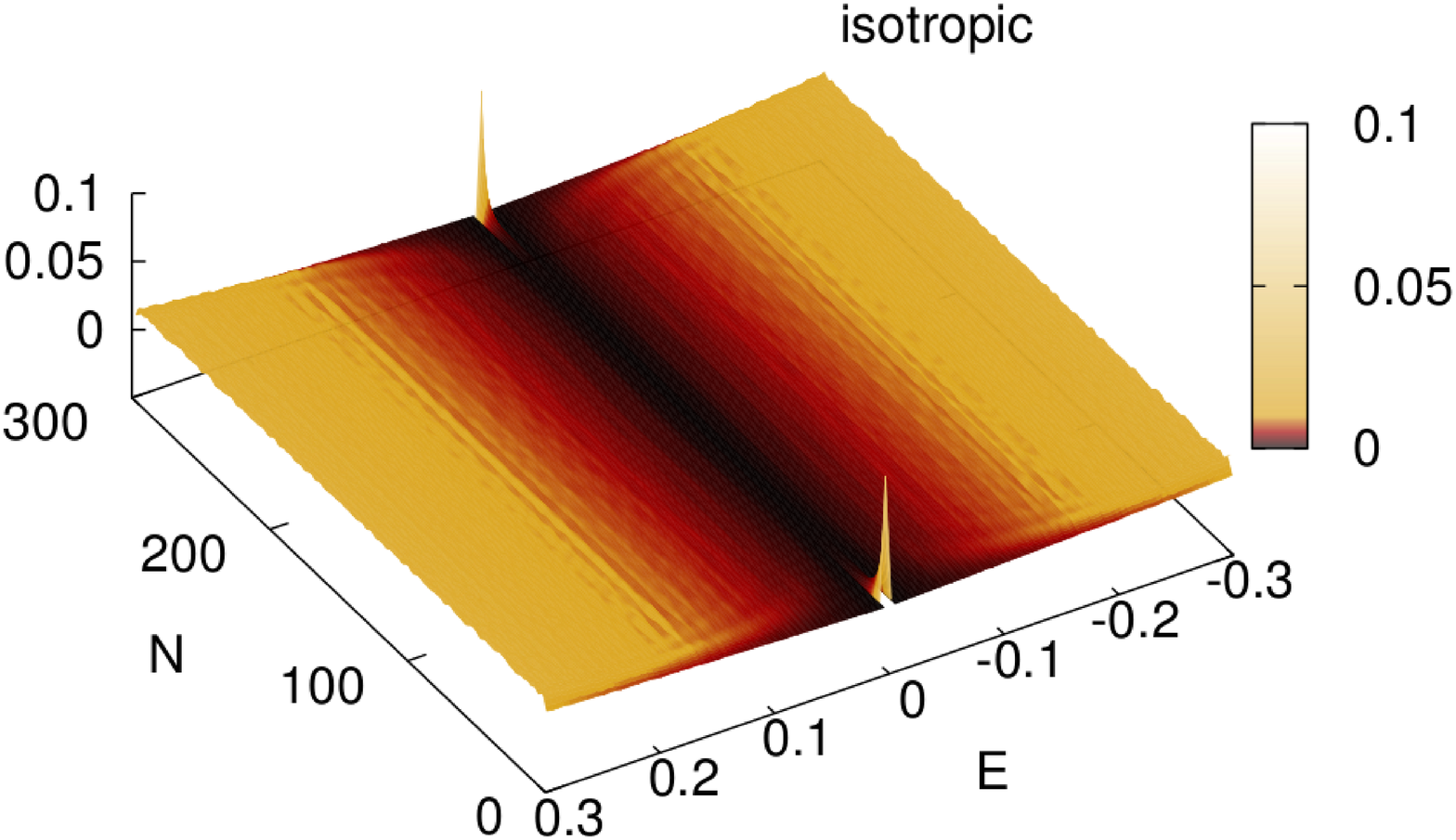}
\caption{(Color online) The low energy spectrum (left) and the  local density of states (LDOS) (right) for isotropic graphene sheet which has infinite length in $x$ direction and finite width of $N=300$ unit cells in $y$ direction. The zero energy edge states exist between two Dirac points and contribute peaks in the LDOS.}
\label{pristine}
\end{figure}

\section{Numerical diagonalization of non-interacting Hamiltonian in strained finite size graphene, subject to magnetic fields}

We consider a single layer graphene with infinite length in $x$ direction and finite width of $N$ unit cells in $y$ direction as shown in Fig. \ref{sketch}. The tight-binding Hamiltonian for spinless fermions (for simplicity) with only nearest-neighbor (NN) hopping reads as
\begin{equation}
H = -\sum_{\langle i,j \rangle}( t_{ij} a_i^{\dagger} b_j + H.c),
\end{equation}
where $a_i$ and $b_j$ are the fermionic annihilation operators on sites of two inter-penetrating triangular sub-lattices. $\langle i,j \rangle$ stands for the summation over three NN sites. In pristine graphene, the hopping amplitudes along three directions are equal. The above Hamiltonian can be represented by a set of Harpers equation, due to the translational symmetry along the x-direction. One can then easily obtain the energy momentum relation numerically. The energy spectrum and the local density of states (LDOS) for isotropic zigzag edge is shown in Fig. \ref{pristine}. The Dirac points are located at $k_x = 2\pi/3$ and $k_x = 4\pi/3$. In the vicinity of the Dirac points the dispersion is linear. The edge states at $E=0$ arise due to charge accumulation on two different sub-lattices, causing peaks in LDOS, which possibly can be detected in STM measurement.  
\begin{figure}
\includegraphics[width=4cm]{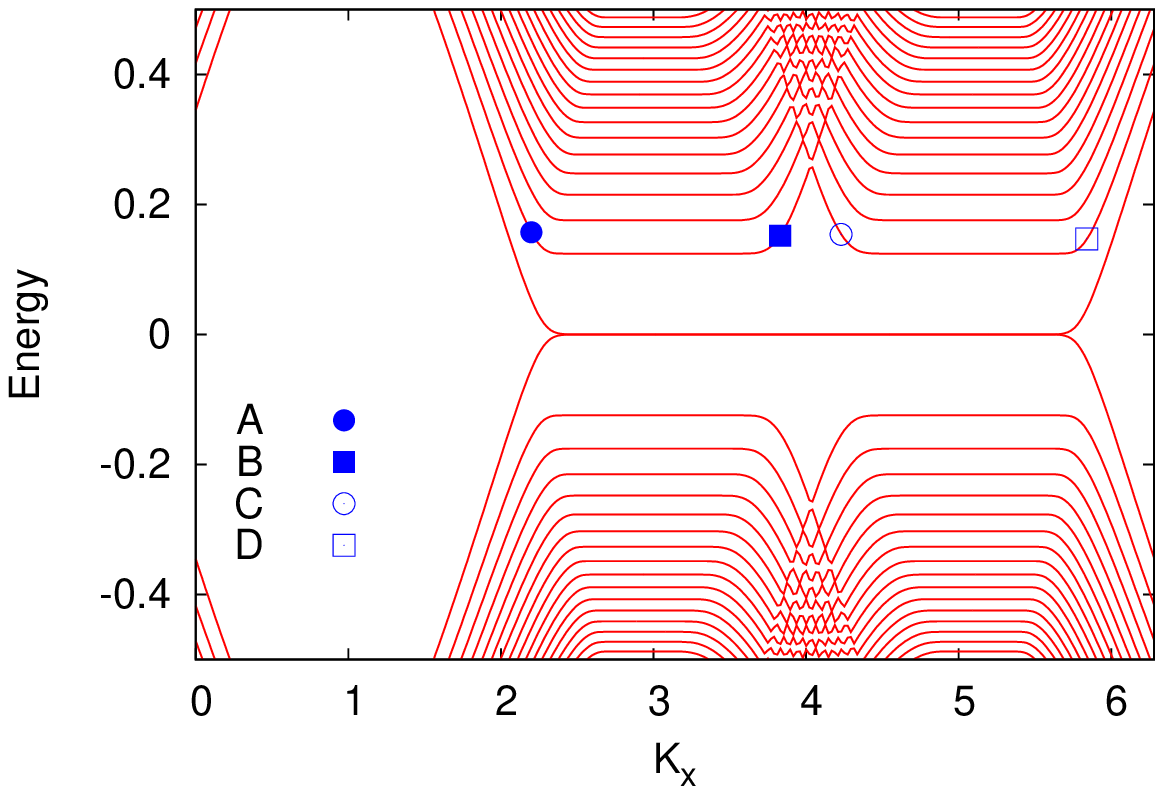}
\includegraphics[width=4cm]{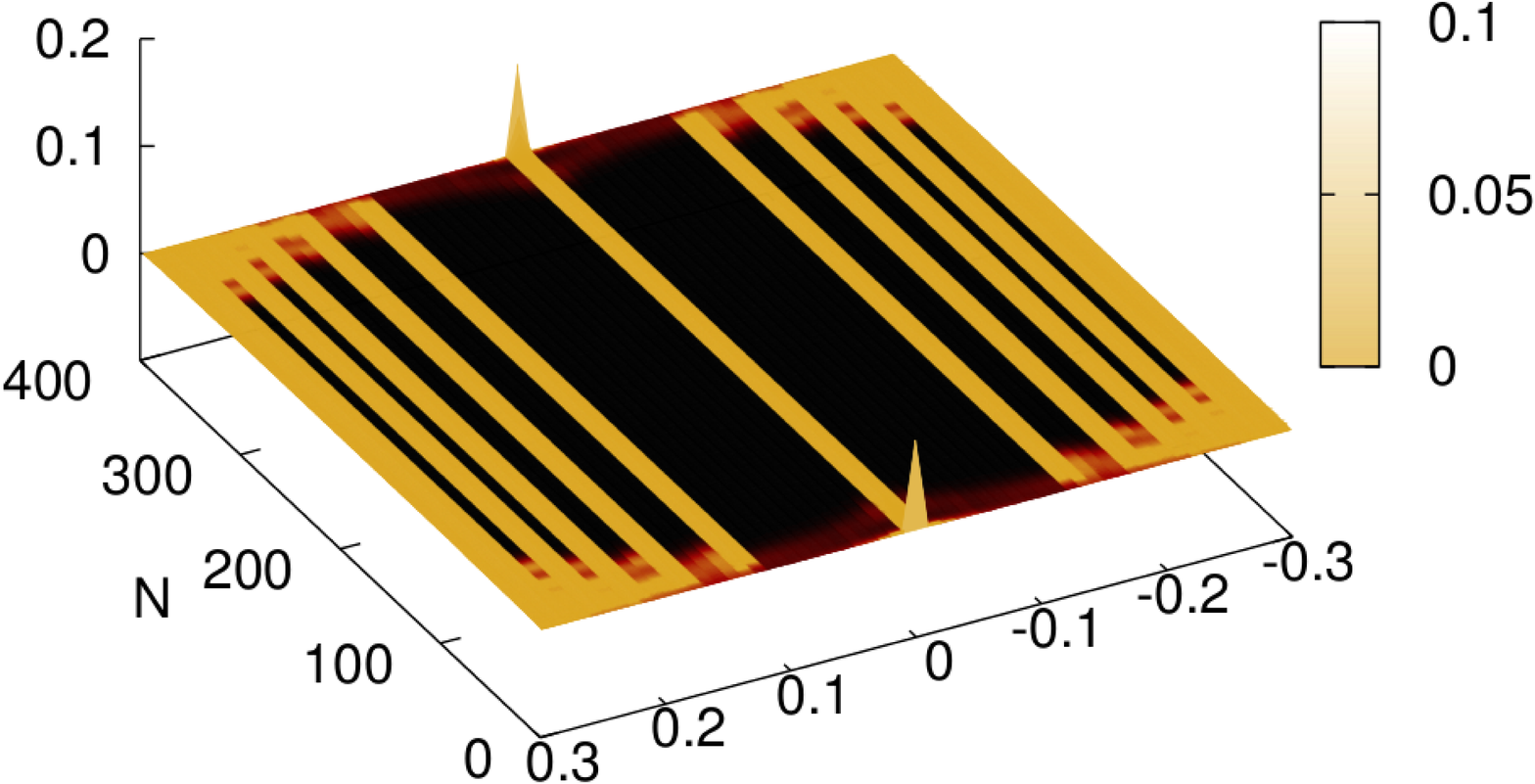}
\includegraphics[width=8cm]{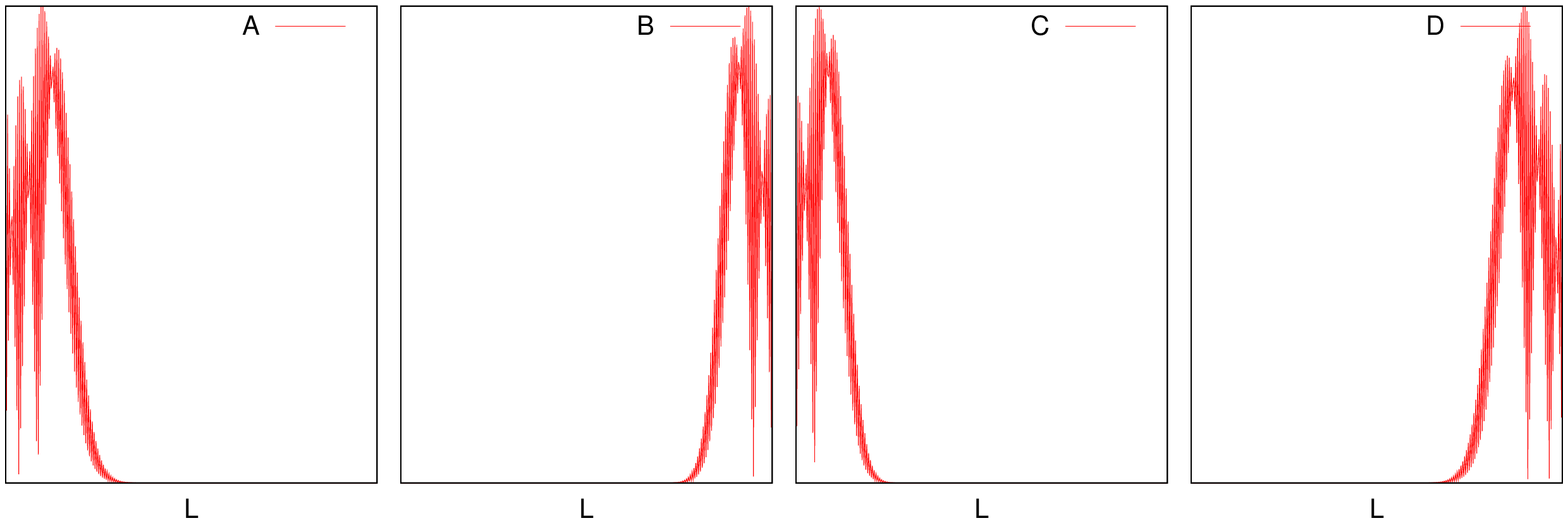}
\caption{\label{Bfield}(Color online) The energy spectrum, LDOS and wavefunction in a real magnetic field. In this case, the Landau levels are clearly developed. The Fermi energy crosses the first Landau level at four points $A, B, C,$ and $D$, in which
A, B and C, D belong to two different valleys. }
\end{figure}

The orbital effect of the real magnetic field can be captured by introducing Peierls phase in the NN hopping amplitudes, $t_{ij} \rightarrow t_{ij} e^{2 \pi i \phi_{ij}}$. $\phi_{ij}$ is related to the magnetic vector potential as 
\begin{equation}
\phi_{ij} = \frac{e}{hc}\int_i^j \vec{A}\cdot d\vec{l}.
\end{equation}
In the presence of magnetic field, the linear dispersions around the Dirac points quenche into sets of Landau levels at well separated energies, as shown in Fig. \ref{Bfield}. It is evident from this figure that the wave functions living on the same edge share identical chirality. For example, wave-functions on left side of each valleys, A and C, are located on left zigzag edge, whereas those on the right side of each valleys, B and D, are located on right zigzag edge.  
\begin{figure}[htb]
\includegraphics[width=7.5cm]{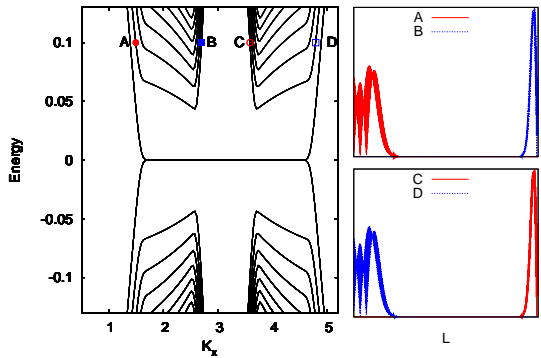}
\caption[] {(Color online) The energy spectrum (left) and the wave functions (right) in a strain field. The hopping, perpendicular to zigzag edge, ranges from t - W/2 to t + W/2 with W = 1.0 from one edge to the other. Here we set $t=1$. Wave-functions localized on one edge, live on opposite side at two valley, explicitly, A(B) and D (C) are localized on left (right) edge. Henceforth, the edge states near two valleys have different chirality. In the presence of real magnetic fields, situation is reversed, i.e., wave-functions on left (A,C) (right (B,D)) side of each valleys are located on left (right) edge.}\label{strain}
\end{figure}

One can introduce the axial magnetic field (b) by locally modifying the hopping amplitudes. One particular way of such deformation is the following. We here modify the hopping only along one of the three bonds, oriented orthogonal to the zigzag edge. It smoothly varies from $t-\frac{W}{2}$ to $t+\frac{W}{2}$, where $W \leq t$, from one edge of the system to the other, separated by $300$ unit cells. With such simple deformation, one ends up with slightly inhomogeneous Fermi velocities and thus LL energies. Nevertheless, we can immediately see from Fig. \ref{strain} that chirality of the edge modes are opposite at each edge in the presence of axial/pseudo magnetic field \cite{chang2}, as we discussed in the manuscript.  
\begin{figure}
\includegraphics[width=7.5cm]{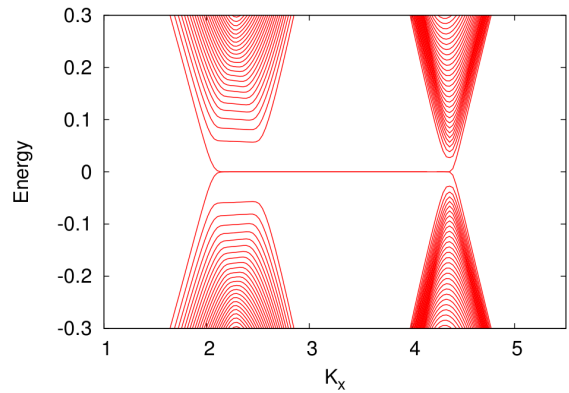}
\caption{\label{Beqb}(Color online) The energy spectrum when B $\approx$ b. The Landau levels are only developed in one valley and the Dirac cone is recovered at the other one.
}
\end{figure}

Upon tuning the relative strength of the fields, one can achieve an interesting limit, $b \approx B$. Only one of the Dirac points is then exposed to finite magnetic field, $2 B$ or $2 b$. Therefore, the energy spectrum near that Dirac point is composed of Landau levels. On the other hand, the system remains semi-metallic in the vicinity of the other Dirac point \cite{tony}. This scenario is depicted in Fig. \ref{Beqb}.

\section{Dirac Landau levels with symmetry breaking terms}

Let us now compute the LL spectrum of the emergent single particle Hamiltonian, containing various symmetry breaking terms. For our purpose, we keep both the anti-ferromagnet ($\vec{N}$) and the quantum spin Hall ($\vec{C}$) orders besides the free Dirac Hamiltonian with real and axial vector potentials. The resulting Hamiltonian reads as 
\begin{eqnarray}
H_{HF}&=& I_2 \otimes i\gamma_0 \gamma_i \left( \hat{q}_i -A_i -i \gamma_{3} \gamma_5 a_i \right) -\left( \vec{N}\cdot \vec{\sigma} \right) \otimes \gamma_0 \nonumber \\
&-& \left(\vec{C} \cdot \vec{\sigma} \right) \otimes i\gamma_1 \gamma_2 
+ \lambda_z \left( \sigma_3 \otimes I_4 \right),
\end{eqnarray}
if one wishes to keep the Zeeman coupling ($\lambda_z$) as well. However, in comparison to the broken symmetry order parameters, the Zeeman term is much smaller and one can neglect it. Then all spin projections are equivalent and we choose $\vec{N}=(N,0,0)$ and $\vec{C}=(C,0,0)$ for the ease of calculation. Even in the presence of symmetry breaking terms two inequivalent valleys are decoupled and one can cast $H_{HF}$ is block diagonal form $H_+ \oplus H_-$, where 
\begin{eqnarray}
H_{\pm} &=& \pm I_2 \otimes \sigma_1 (\hat{q}_1 \mp a_1 -A_1) + I_2 \otimes \sigma_2 (\hat{q}_2 \mp a_2 -A_2) \nonumber \\
&-& \left( \sigma_3 \otimes \sigma_3 \right) N \mp \left( \sigma_3 \otimes \sigma_3 \right) C,
\end{eqnarray}
after exchanging the second and third $2 \times 2$ blocks. Using the unitary rotations $U_1$ and $U_2$, defined in sec. I, one can cast $H_{\pm}$ as 
\begin{equation}
H_\pm \rightarrow i \gamma_0 \gamma_i \left( \hat{q}_i \mp a_i -A_i \right) - \gamma_0 \left( N \pm C \right).
\end{equation}
Therefore the spectrum of $H_{HF}$ is composed of two sets of inequivalent LLs, at energies $\pm \sqrt{2 n (b \pm B)+ \left( N \pm C \right)^2}$, for $n \geq 1$ and $\pm |N \pm C|$ when $n=0$.

\section{ Gap equations}

The ground state energy per unit area  at half-filling in the presence of two symmetry breaking order parameters reads as 
\begin{equation}
E [N,C]=\frac{N^2}{4 g_a}+\frac{C^2}{4 g_c}+ E_0 [N,C],
\end{equation}
as shown in the main text. $E_0[N,C]$ is the ground state energy of the effective single particle Hamiltonian $H_{HF}$. At Half-filling, the ground state of $H_{HF}$ has all the states at negative energies filled, while those at positive energies at completely empty, yielding 
\begin{eqnarray}
E_0[N,C]&=&\frac{1}{4 \pi}\sum_{n \geq 1}\bigg( (b+B)e_{n,+}+(b-B) e_{n,-} \bigg) \nonumber \\
&+& \frac{1}{4 \pi} \bigg( (b+B) (N+C) + (b-B) |N-C| \bigg), \nonumber \\
\end{eqnarray}
where $e_{n,\sigma}=\sqrt{2 n (b + \sigma B)+ \left( N +\sigma C \right)^2}$, for $\sigma=\pm 1$. In the main text of the paper, we argued that unless $g_a \equiv g_c$, the anti-ferromagnet and the spin Hall order cannot coexist. For $g_a > g_c$, i.e when the on-site repulsion (U) is stronger than the next neighbor one ($V_2$), at half-filling the ground state configuration is with $N \neq 0$, while $C \equiv 0$. Minimizing the ground state energy, $E[N,0]$, with respect to $N$, then yields the gap equation at half filling
\begin{eqnarray}
\frac{1}{4 g}&=&\frac{1}{4 \pi^{3/2}} \int^{\infty}_0 \frac{ds}{s^{3/2}} \; e^{-s m^2} \sum_{\sigma=\pm 1} \left(b +\sigma B \right) 
\frac{2 s}{e^{2 s (b +\sigma B)}-1} \nonumber \\
&+& \frac{b}{2 \pi m},
\label{gapzero}
\end{eqnarray}   
where $g \equiv g_a$. The first term on the R. H. S. counts the contribution of all the filled LLs with $n \geq 1$, and it is ultra-violet divergent, while the second term on the R. H. S. arises only from the ZLL. However, the divergence can be regulated by defining the zero field critical interaction for ordering as 
\begin{equation}
\frac{1}{g_*}= \frac{2 \Lambda}{\sqrt{\pi}} \int^\infty_0 \frac{d s }{s^{3/2}},
\end{equation} 
after $\pi/g \rightarrow 1/g$, so that the physical gap remains cut-off ($\Lambda$) independent. The cut-off ($\Lambda$) defines the range of energy, over which the dispersion is approximately linear. Defining the strength of the interaction as 
\begin{equation}
\delta= \frac{1}{\Lambda g} - \frac{1}{\Lambda g_*}, 
\end{equation}
where $\delta$ measures the deviation of the interaction from the zero field critical interaction for insulation, one can cast the gap equation as 
\begin{equation}
x \left( 1+ \frac{1}{q}\right)= \frac{\delta}{m} + \frac{1}{\sqrt{\pi}} \left( 1 + \frac{1}{\sqrt{q}}\right) f(x).
\end{equation} 
Here we introduced new variables, $x=\frac{b+B}{m^2}$, and $q=\frac{b+B}{b-B}$ and the function $f(x)$ reads as 
\begin{equation}
f(x)=\frac{m \; \sqrt{x}}{\sqrt{\pi}} \left( 1+\frac{1}{\sqrt{q}}\right) \;
\int^\infty_0 \frac{d s }{s^{3/2}} \; \bigg[ \frac{2 s e^{-s/x}}{e^{2 s}-1} - 1 \bigg].
\end{equation}

Similarly, one can arrive at the gap equation for the $\sigma_{xy}=\pm e^2/h$ Hall state, which on the other hand reads as 
\begin{eqnarray}
\frac{1}{4 g}&=&\frac{1}{4 \pi^{3/2}} \int^{\infty}_0 \frac{ds}{s^{3/2}} \; e^{-s m^2} \sum_{\sigma=\pm 1} \left(b +\sigma B \right) 
\frac{2 s}{e^{2 s (b +\sigma B)}-1} \nonumber \\
&+&\frac{b+B}{4 m \pi},
\label{gapone}
\end{eqnarray}
where $g \equiv g_c$, assuming $N>C$ or $g_a>g_c$. Here the contribution from the filled LLs (first term on R. H.S.) is identical to that in Eq. (\ref{gapzero}), but the ZLL contribution is different from that in Eq. (\ref{gapzero}), since only fewer states contributes to the ground state energy when the chemical potential is tuned close to the spin polarized gap (N). After performing similar exercise, one can finds the gap equation to be
\begin{equation}
x = \frac{\delta}{m} + \frac{1}{\sqrt{\pi}} \left( 1 + \frac{1}{\sqrt{q}}\right) f(x).
\end{equation}

For $\delta=0 \;(g=g_*)$ and $b=300$ T, solutions of the gap equations are at $x=x_0=5.0537$ and $17.9226$ for $B=0$ T, respectively. For $B=50$ T, the solutions are at $x_0=5.741$ and $15.378$.\cite{bitanscalingall} For a finite $\delta >0 \; (g<g_*)$, the solutions can only be found at $x>x_0$. Hence, for realistic strength of two fields, one can safely expand $f(x)$ for large argument of $x$, leaving us with simple \emph{algebraic} equations for the interaction induced activation gap for $\sigma_{xy}=0,\pm e^2/h$ Hall state, read as 
\begin{eqnarray}
&&x \left( 1+ \frac{1}{q}\right)-\frac{\delta}{m} \nonumber \\
&-& \left(1+\frac{1}{\sqrt{q}} \right) \: \left( \sqrt{x} u + \frac{v}{\sqrt{x}} + {\cal O} (x^{-3/2}) \right)=0,
\end{eqnarray}    
\begin{eqnarray}
x -\frac{\delta}{m}- \left(1+\frac{1}{\sqrt{q}} \right) \: \left( \sqrt{x} u + \frac{v}{\sqrt{x}} + {\cal O} (x^{-3/2}) \right)=0,  \nonumber \\
\end{eqnarray}    
respectively. The coefficients are $u=2.0652$ and $v=0.924$. The numerical solutions of the gaps are shown in the main part of the paper, for a set of realistic strengths of interactions $(\delta >0)$ in graphene.


\begin{thebibliography}{99}
\bibitem{graphenechirality}  With only $B$, particle and hole LLs have opposite chirality, hence $\nu \epsilon \pm {\cal Z}$.
\end{thebibliography}
\end{document}